# An Efficient and Robust Method for Chest X-Ray Rib Suppression that Improves Pulmonary Abnormality Diagnosis

Authors
Di Xu: DiXu@mednet.ucla.edu
Qifan Xu: QifanXu@mednet.ucla.edu
Kevin Nhieu: kknhieu7@g.ucla.edu
Dan Ruan: DanRaun@mednet.ucla.edu
Ke Sheng: KSheng@mednet.ucla.edu (corresponding author)

## Abstract

Suppression of thoracic bone shadows on chest X-rays (CXRs) has been indicated to improve the diagnosis of pulmonary disease. Previous approaches can be categorized as unsupervised physical and supervised deep learning models. Nevertheless, with physical models able to preserve morphological details but at the cost of extremely long processing time, existing DL methods face challenges of gathering sufficient/qualitative ground truth (GT) for robust training, thus leading to failure in maintaining clinically acceptable false positive rates. We hereby propose a generalizable yet efficient workflow of two stages: (1) training pairs generation with GT bone shadows eliminated in by a physical model in spatially transformed gradient fields. (2) fully supervised image denoising network training on stage-one datasets for fast rib removal on incoming CXRs. For step two, we designed a densely connected network called SADXNet, combined with peak signal to noise ratio and multi-scale structure similarity index measure objective minimization to suppress bony structures. The SADXNet organizes spatial filters in U shape (e.g., X=7; filters = 16, 64, 256, 512, 256, 64, 16) and preserves the feature map dimension throughout the network flow. Visually, SADXNet can suppress the rib edge and that near the lung wall/vertebra without jeopardizing the vessel/abnormality conspicuity. Quantitively, it achieves RMSE of ~0 during testing with one prediction taking <1s. Downstream tasks including lung nodule detection as well as common lung disease classification and localization are used to evaluate our proposed rib suppression mechanism. We observed 3.23% and 6.62% area under the curve (AUC) increase as well as 203 and 385 absolute false positive decrease for lung nodule detection and common lung disease localization, separately.

1. Introduction

Respiratory diseases are among the major causes of morbidity and mortality globally and the prevalence of pulmonary diseases have been continuously increasing for a long period of time [1, 2]. Chest X-ray (CXR) imaging, considering its affordability and accessibility, is currently most widely used for health monitoring and pre-screening of thoracic abnormalities including lung carcinoma, tuberculosis, pneumonia, pneumothorax as well as pulmonary emphysema [3]. Particularly, CXR plays an essential role for patients in intensive care units (ICUs) for its portability and mobility [4]. Nevertheless, the detection task of lung diseases on CXR is challenging because of increased image noise resulting from superimposed anatomies. The high-contrast bony structures are one of the major noise contributors since a signal of interest in CXR can be partially or completely obscured or overshadowed by its surrounding ribs. Therefore, suppression of thoracic bone shadows on CXR is extremely desirable to improve the visibility of soft tissue.

In recent years, clinical evidences indicates that rib-removed CXR can improve the diagnosis of various pulmonary abnormalities[5, 6]. In a post-processing setup, previous approaches can be summarized as: (1) unsupervised physical models [3, 7] (2) deep learning (DL) models with supervision generated from dual-energy (DE) CXRs or domain adapted digitally reconstructed radiograph (DRR) [8, 9].

Physical models utilize single-energy (SE) CXR, reconstruct the rib structures via assuming various hypothesis on the pixel intensity distribution of ribs, and obtain the bone-suppressed results via subtracting the generated rib shadows from original CXRs. For example, von Berg et al. proposed to transfer chest bones into the so-called ST-space where the contour of ribs appears as a straight line. Then partial derivative computation, smoothing, and reintegration are conducted along the rib contours up to the centerline. The idea of derivative smoothing is to remove all other signals besides the bone contour from image, then reintegration to reconstruct the bone signals, and lastly subtract the reconstructed bone signals from the original image to generate the rib suppressed CXR [7]. This type of methods typically can well preserve morphological details per bone suppression, yet it is at the cost of not only requiring manual annotation of rib masks for each CXR, but also case-by-case parameter fine-tuning as well as impractically long prediction time, which usually takes around half to an hour to generate a single radiograph [5, 7, 10–12].

On the contrary, DL methods are time-efficient in image prediction, but its performance substantially depends on the quality of training supervision and quantity of training pairs [13]. Apart from diverse network schemes for superior representation learning, the core disparities of existing DL-based approaches are in the modality of training image and method of training pair generation. For image modality, most of the previous models are trained on either DE CXR or DRR. Regarding training pair generation, since DE chest radiograph is formed by using two XR exposures at distinct energy levels, which can later be decomposed into "soft-tissue-only" and "bone-only" images [14, 15]. Learning models trained on DE CXR forms image pairs with unprocessed images as input and decomposed "soft-tissue-only" images as ground truth (GT) [3, 16]. Nevertheless, DE CXR based frameworks commonly suffers from insufficient training due to limited amount of DE CXR available [17]. Moreover, since bony structures are separable in computed tomography (CT) domain, DRR based learning methods attempt to project training pairs from CT images with and without ribs [18]. Nonetheless, even though DRR looks like CXR, they have disparate image contrast due to different simulation and post-processing setup [19]. Because sampling perfectly paired CXR and DRR is challenging in practice, current domain adaptation techniques are restrictive to unpaired GAN-based models trained with input and GT of unpaired DRR and CXR with ribcage [19, 20]. The hurdle of this design is that the adaptation model is solely trained to understand the mapping from DRR with ribs to CXR with ribs, it will, to a certain extent, add back the bony structure into DRR without ribs while adapting it into CXR domain. To this end, we conclude that to sufficiently train a "well-supervised" DL network for the task of CXR rib removal, an adequate dataset with dedicated GT is in tremendous need.

To sum up, physical models can predict qualitative bone suppressed images yet require long running time and labor-intensive manual rib annotations and parameter tuning. On the contrary, DL methods can greatly expedite prediction time and require no rib annotation as well as image-wise parameter tuning yet lack large image pairs and with high-quality GT supervision for model training. Therefore, we believe that utilizing a physical model to prepare a qualitative benchmark dataset first and then develop various DL frameworks based on the benchmark dataset is a potential solution for fast and generalizable rib suppression on CXR scans.

On that account, we hereby introduce a new benchmark dataset named FX-RRCXR for DL-based CXR rib suppression. The FX-RRCXR is generated on the basis of VinDr-RibCXR dataset [21] using the physical model called ST-smoothing improved from von Berg et al [7]. Next, we propose a novel full supervised image denoising network – SADXNet – trained on FX-RRCXR for fast rib suppression of incoming CXRs. Lastly, we validate the impact of CXR rib suppression in object detection tasks including lung nodule detections using NODE21 [22] dataset and general lung disease classification and localization using ChestX-ray14 [23] datasets. All the detection tasks are run on Mask R-CNN [24] framework.

## 2. Method

In this section, we will introduce first the ST-smoothing method, next the FX-RRCXR dataset, then the denoising network SADXNet, and lastly the NODE21 and ChestX-ray14 datasets. Details will be elaborated as follows.

### 2.1 ST-smoothing

The fundamental assumption of ST-smoothing algorithm is defined in $Assumption$ 1. As shown in S0 of **Figure 1**, if the distances to centerline of $p_1$ and $p_2$ are equal, we will assume that these two points are on the same contour and their rib intensities are $p_1 = p_2$. The overall workflow of ST-smoothing algorithm is presented in **Figure 1**, which will be elaborated in the following subsections.

***Assumption* 1**: *the pixel intensities along one continuous contour of a rib are theoretically identical.*

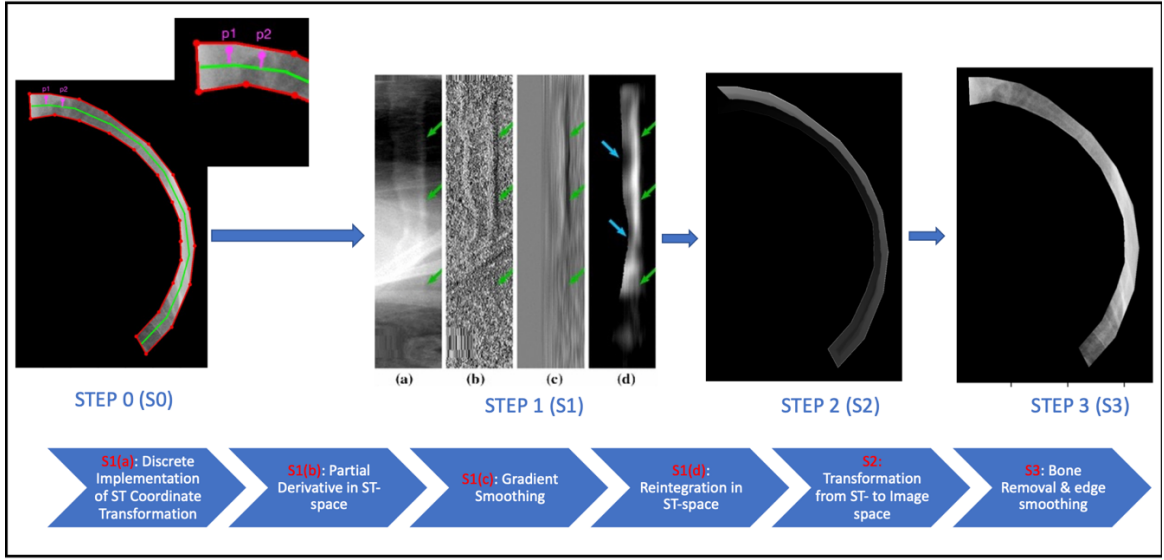

**Figure 1.** Major steps of the ST-smoothing algorithm applied to the fifth rib of CXR with image visualization in the upper and pipeline description in the bottom section. Explanation for the purpose of each image is labeled in red on flow chart.

### 2.1.1 From Image Space to ST-coordinate System

$$T_C: (x, y) \mapsto (s, t) \quad (1)$$
$$T_C^{-1}: (s, t) \mapsto (x, y) \quad (2)$$

The current sub-section illustrates the step from S0 to S1(a) in **Figure 1**. ST-transformation $T_C$ is a domain transformation used to generate a specific representation of a part of the image defined by the given closed cyclic contour $C: \gamma(t), t \in [0, C_{len})$ with $\gamma(0) = \gamma(C_{len})$. $T_C$ will be defined by its inverse in equation (2), which is given by

$$T_C^{-1}(s, t) = \gamma(t) + s \cdot \frac{\gamma'(t)}{|\gamma'(t)|}^{\perp} \quad (3)$$

where $\frac{\gamma'(t)}{|\gamma'(t)|}^{\perp}$ is the contour norm at contour point $\gamma(t)$. **Figure 2 (a)** shows the transformation of contour C [7].

To reduce the computation burden, $\gamma(t)$ is considered as piecewise linear closed contour as shown in **Figure 1** S0 and then the implementation of equation (1) becomes discrete instead of continuous. **Figure 2 (b)** illustrates the implementation process, where $s$ and $t$ are formulated as

$$s = ||\overrightarrow{QQ'}||$$
$$t = t_{prev} + ||\overrightarrow{QP_i'}|| \cdot \frac{||P_i P_{i+1}||}{||\overrightarrow{P_i' P_{i+1}'}||} \quad (5)$$

where $t_{prev}$ is the summed length of all previous edges. The pixel intensity in the ST-domain is given by

$$I_{st_C}(s, t) = I(T_C^{-1}(s, t)) \quad (6)$$
$$(\hat{s}, \hat{t}) = T_C(T_C^{-1}(s, t)) \quad (7)$$

Assume that a valid $(s,t)$ always has a position $(\hat{s}, \hat{t})$ on the other side of the bone centerline $c(t)$ defined in equation (7). $c(t)$ could be obtained from

$$c(t) = \max_{S} \forall\, (\hat{s}, \hat{t}) \neq (s,t) \tag{8}$$

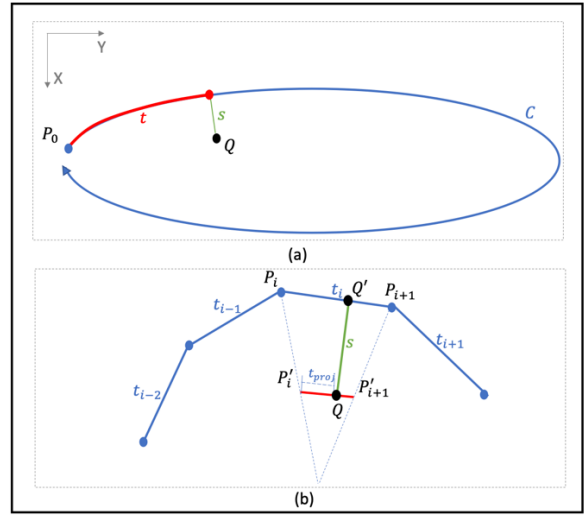

**Figure 2**. Visualization of the $ST$ - coordinate system. (a) Point $Q$ transformation from $XY$- into $ST$-domain regarding contour $C$. (b) Discrete sampling in ST-coordinate system with respect to point $Q$.

*2.1.2 Rib Extraction via Partial Derivatives Smoothing in ST-space*
Under this subsection, we will focus on explanation from step S1(a) to S2 in **Figure 1**, which includes discrete partial derivative calculation in ST-space, gradient smoothing, gradient reintegration, and lastly transformation from ST- into image-space.

2.1.2.1 Discrete Partial Derivative in ST-space
As shown in equation (9). The first order partial derivative is calculated along the $s$ axis in a discrete form to speed up the overall computation. Under $Assumption\ 1$, the definition of $I_{d_C}(s,t)$ represents the gradient orthogonal to $t$-axis of rib and can be interpreted as any structure oriented along the axis $t$ having no contribution to the gradients of bone.

$$I_{d_C}(s,t) = \partial s I_{st_C}(s,t) = I(s,t) - I(s-1,t) \tag{9}$$

2.1.2.2 Smoothing, Reintegration, and Transformation back to XY-domain
Improved from von Berg et al.[7], two types of smoothing are implemented including gaussian smoothing ($\mathcal{G}_{k_t}$) along the $t$-axis on $I_{d_C}$ and centerline smoothing ($\mathcal{C}$) along the $s$-axis on reintegrated $I_{r_C}$. First and foremost, since we hypothesis that the signals along $t$-axis of $I_{d_C}$ has nothing to do with ribs, large gaussian kernel $k_t$ can be used in this case to smooth out the signals from soft tissue and leave only signal of interest from ribs in $\mathcal{G}_{k_t}(I_{d_C})$. Noteworthily, $k_t$ is a hyper-parameter which needs image base tuning. Next, after excluding the soft tissue signal via

$\mathcal{G}_{k_t}(I_{d_C})$, we conducted reintegration towards the smoothed partial gradients to recover the bone signal $I_{r_C}(s,t)$ in ST-domain, as shown in equation (10).

$$I_{r_C}(s,t) = \int_{s-1}^{s} \mathcal{G}_{k_t}(I_{d_C}(s,t)) + \mathcal{G}_{k_t}(I_{d_C}(s-1,t)) \tag{10}$$

Lastly, for the fact that $T_C$ has a singularity at the centerline $c(t)$, an artificial edge can be observed along $c(t)$ as shown in **Figure 1** S2 after reintegration of $I_{r_C}$. Therefore, a K-nearest-neighbor (KNN) based centerline processing function is applied along the $s$-axis of $I_{r_C}$ to smooth out the edge (shown in equation (11)).

$$\mathcal{C}\left(I_{r_C}(s_i, t_j)\right) = \begin{cases} I_{r_C}(s_i, t_j), & if\ \dfrac{I_{r_C}(s_i, t_j)}{I_{r_C}(s_{i-1}, t_j)} > \tau \\ \dfrac{\sum_{m=i-k+1}^{i} I_{r_C}(s_m, t_j)}{k}, & otherwise \end{cases} \tag{11}$$

Where $\tau$ and $k$ are two hyper-parameters requires patient-wise tuning and represents threshold for conducting KNN average and number of neighbors used, respectively.

After all the above steps, we will transfer the rib intensity $\mathcal{C}(I_{r_C})$ from ST- back into image-space under the condition of equation (12) to exclude the potential negative values, which is uninterpretable in image space.

$$I_{bone_C} = \max\left(\mathcal{C}(I_{r_C}), 0\right) \tag{12}$$

### 2.1.3 Rib Removal and border blending

We will focus on the explanation of **Figure 1** S3 here. The rudimentary soft tissue CXR $I_{soft_C}$ is acquired via subtracting $I_{bone_C}$ from the raw CXR as

$$I_{soft_C}(x,y) = I(x,y) - I_{bone_C}(x,y) \tag{13}$$

Lastly, to improve the continuity between rib boundary $r_b$ and its surrounding soft tissue, a KNN border smoothing algorithm is applied to $r_b$

$$r_b'(x,y) = \frac{\sum_{i=1}^{k} r_{b,i}(x,y)}{k} \tag{14}$$

Where the $r_b$ is defined under ST-space and shown equation (15). $s_b$ in equation (15) and $k$ in equation (14) are two hyper-parameter requires value search.

$$\begin{aligned} I_{soft_C}(s,t)\ in\ r_b(s,t), & \quad if\ s \leq s_b \\ r_b(x,y) = T_C^{-1}(r_b(s,t)) & \end{aligned} \tag{15}$$

And generating a complete soft tissue CXR requires iteratively repeating section 2.1.1 to 2.1.4 for every rib annotation in the ribcage mask.

## 2.2 Data Cohort
### 2.2.1 VinDr-RibCXR Dataset

VinDr-RibCXR dataset is selected for processing of FX-RRCXR dataset using ST-smoothing algorithm. VinDr-RibCXR is a benchmark dataset prepared for automatic rib segmentation and labelling of SECXR scans. This dataset contains 245 images with corresponding GT individual rib masks annotated by human expert. Each CXR scan has 20 separate rib annotations in the left and right side of lung. The dataset was pre-split into training and validation sets with 196 scans in training and 49 in validation. We refer audience to Nguyen et al. [21] for more details.

### 2.2.2 FX-RRCXR Dataset Preparation

We sent all the 245 images along with their rib masks into ST-smoothing algorithm for bone shadow suppression. The hyper-parameters required by ST-smoothing are tuned image base using random grid search algorithm. Excluding the process for hyper-parameter searching, predicting a single soft tissue image using TS-smoothing took from around 40 to up to 70 minutes. The time variation is determined by the number of pixels within each ribcage mask. We organize the original CXR as input and its corresponding soft tissue scan as GT while preparing image pairs and kept the same training and validation set split as VinDr-RibCXR dataset.

### 2.2.3 Augmentation for SADXNet Training

For more robust training, augmentation including random resizing in the scale of [0.8-2, step=0.2], random rotation in the angle of (0°, 180°), random gaussian blur with kernel size of (5, 5) and sigma range of (0.1, 0.3) are implemented towards the training set of FX-RRCXR dataset.

### 2.3 SADXNet

### 2.3.1 Network Structure

Since rib signals can be treated as noise on soft tissue CXRs, we herewith propose to design an image denoising network named SADXNet for training of the FX-RRCXR dataset. Inspired by the architecture of DenseNet [25], SADXNet is designed to be densely connected, which connects each layer to every other layer in a feed-forward fashion as shown in **Figure 3**. In specific, for each layer in SADXNet, the feature maps of all preceding layers and its own feature map are used as inputs into all subsequent layers. The advantage of densely connecting the SADXNet is to better alleviate the vanishing-gradient problem, strength feature propagation, and encourage feature reuse [25]. We will elaborate on the composition of SADXNet as following.

**Dense connectivity**. Having **Figure 3** illustrating the layout of densely connected SADXNet schematically, we will also define the mechanism of each layer in equation (16).

$$x_l = N_l([f_{c_l}(x_0), f_{c_l}(x_1), \ldots, f_{c_l}(x_{l-2}), x_{l-1}]) \qquad (16)$$

Where $[x_0, x_1, \ldots, x_{l-1}]$ refers to the channel-wise concatenation of the feature maps produced from layer 0 to $l-1$ and $f_{c_l}(\cdot)$ composes of a $1 \times 1$ convolution (Conv) to unify the number of feature channels of $x_0$ to $x_{l-2}$. The output channel of $f_{c_l}$ is defined by $\frac{C(x_{l-1})}{L-1}$, where $C$ represents the number of feature channels of $x_{l-1}$. The purpose $f_{c_l}(\cdot)$ is to avoid the concatenated input feature maps to $N_l(\cdot)$ being overly large in the channel dimension and will not fit in the GPU memory.

**Composite function.** We define $N_l(\cdot)$ as a composite function of three consecutive operations: batch normalization (BN), followed by a rectified linear unit (ReLU), and a 3*3 Conv.

**Pooling layers.** Inspired by pioneer study [8], we preserve the height (H) and width (W) dimensions of feature map as the input shape of input image throughout the Conv process without using down- or up-sampling layers which will change the size of feature maps.

**Channel design.** There are 7 densely connected layers in total in the SADXNet structure. The channel of convolutional kernel for each dense layer is designed in an increase-to-decrease setting to mimic the design of fully convolutional network [26] in channel dimension. The purpose of this design is to balance between model complexity (kernel with more channel) and training time (kernel with less channel).

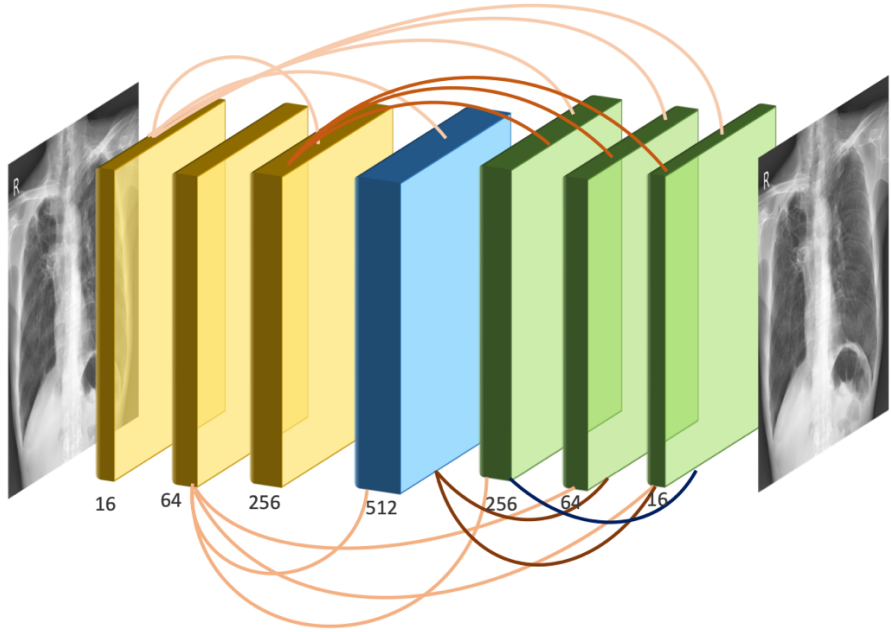

**Figure 3**. The network architecture of SADXNet.

**Loss Function.** The cost function for training SADXNet is designed in the form of a combination of negative peak signal-to-noise ratio (PSNR), multi-scale structure similarity index measure (MS-SSIM) [27], and $L_1$ deviation measurement:

$$L = -\alpha \cdot L_{PSNR} + (1-\alpha) \cdot [\beta \cdot L_{MS-SSIM} + (1-\beta) \cdot L_1] \quad (17)$$

$$L_{PSNR} = log_{10}\left(\frac{MAX_X^2}{1/mn \cdot \sum_{i=0}^{m-1}\sum_{i=0}^{n-1}[x_{ij} - y_{ij}]^2}\right) \quad (18)$$

$$L_{MS-SSIM} = 1 - \frac{(2\mu_x\mu_y + c_1)}{(\mu_x^2 + \mu_y^2 + c_1)} \cdot \prod_{j=1}^{M} \frac{(2\sigma_{x_j y_j} + c_2)}{(\sigma_{x_j}^2 + \sigma_{y_j}^2 + c_2)} \quad (19)$$

$$L_1 = \frac{1}{mn} \cdot \sum_{i=0}^{m-1}\sum_{i=0}^{n-1} ||x_{ij} - y_{ij}||_1 \quad (20)$$

Where $\alpha$ and $\beta$ in equation (17) are hyperparameters set to 0.75 and 0.25, $X$ and $Y$ are model input and target, $MAX_X$ is the maximum possible input value, $[\sigma_{x_j y_j}, ..., \sigma_{x_M y_M}]$ of equation (18) are set to $[0.5, 1.0, 2.0, 4.0, 8.0]$, $c_1 = (k_1 S)^2$ and $c_2 = (k_2 S)^2$ of equation (19) are two variables to stabilize the division with weak denominator having $S$ as the dynamic range of the pixel-values (typically $2^{\#\ bits\ per\ pixel} - 1$) and $(k_1, k_2)$ are constants, and $||\cdot||_1$ denotes the $l_1$ norm.

**Model Training.** SADXNet was implemented in PyTorch and the training was performed on a GPU cluster with $4 \times RTX\ A6000$. For training of SADXNet, we set the maximum number of epochs as 200 and observed that the model converged at around 100 epochs with validation loss reaching the elbow point. Adam optimizer with initial learning rate (LR) of 0.001 and batch size of 1×4 was applied during learning.

3. Results
3.1 Results of Rib Suppression

In this section, we will present and analyze the visual results from ST-smoothing as well as the visual plus quantitative results from SADXNet.

3.1.1 ST-smoothing

**Figure 4** demonstrates a patient with lung nodules in the left lung field and compares the visibility of nodes with and without rib removal using ST-smoothing algorithm. We can tell that in comparison with previous rib-suppression methods which have rib edge residuals or artifacts near lung borders [5, 19], ST-smoothing carefully avoids those two aspects. Additionally, we observe that ST-smoothing can well preserve the shape and morphological details of lung tumors while suppressing the ribcage. Lastly, we found that we are more confident in visually localizing the nodules with rib suppression than without.

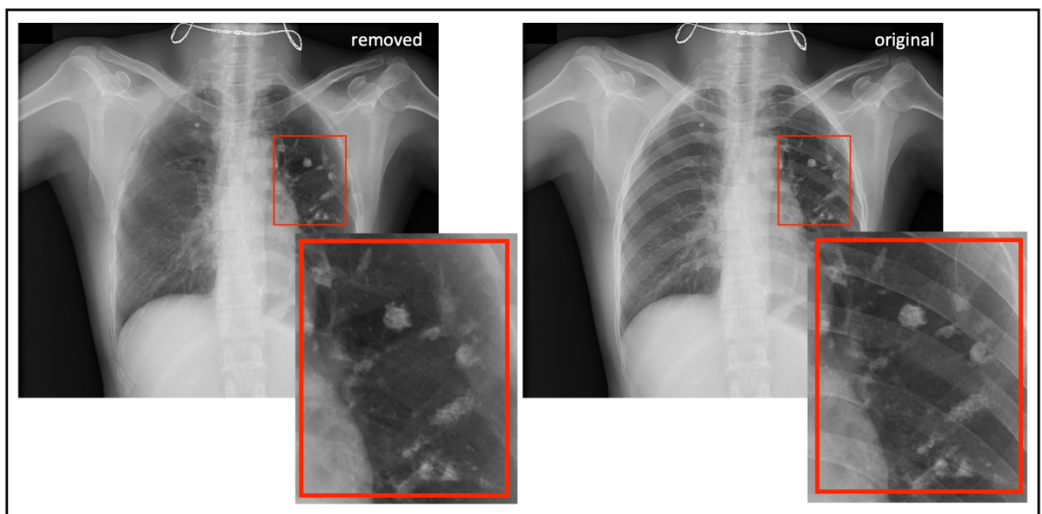

**Figure 4**. A sample case from ST-Smoothing with several lung nodule in the left lung field. The left-hand side shows rib-removed scan and the right-hand side shows the original unprocessed CXR image.

### 3.1.2 Results of SADXNet

**Figure 5** presents the results of two patients predicted by SADXNet. We can notice that the soft tissue scans predicted by SADXNet are visually indistinguishable from their corresponding GTs. Quantitatively, SADXNet achieves almost 0 rooted mean squared error (RMSE) per model evaluation on test set. Finally, compared to ST-smoothing algorithm which takes up to an hour to predict a soft tissue image, the trained SADXNet takes <1 second to make one prediction.

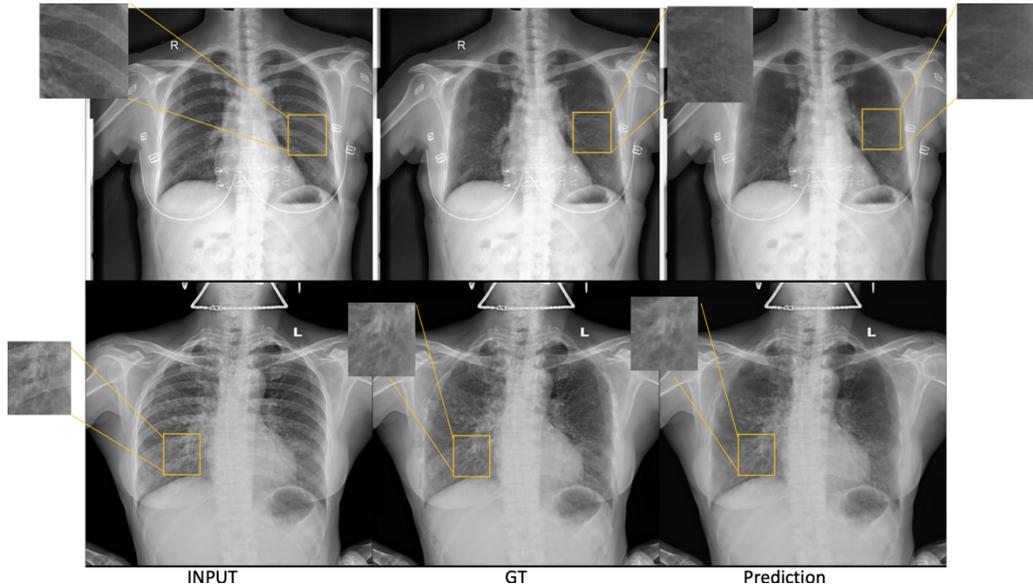

**Figure 5.** Two sample cases from test set to illustrate the results of SADXNet rib removal. Each row represents one patient. The columns from left to right are ordered as the first column showing raw CXR, the second one showing soft tissue image generated from ST-smoothing (GT for SADXNet training supervision), and the third one showing soft-tissue CXRs predicted from trained SADXNet model.

### 3.2 Downstream Task Validation

Ultimately, the value of soft tissue CXRs should be demonstrated by its contribution in downstream applications. We quantify the benefits of using rib suppressed image with two separate experiments including a lung nodule detection task based on NODE21 dataset [22] and a general pulmonary disease classification and localization task based on ChestXRay14 dataset [23]. Both detection tasks are run on Mask R-CNN object detection pipeline with mask head off [24]. We conduct experiments on three different input combinations with all the rib suppressed images used below predicted from SADXNet: 1) only input of raw CXRs. 2) only input of soft-tissue CXRs. 3) mixed input of raw and soft-tissue CXRs. Lastly, area under the curve (AUC), true positive predictions (TP), false positive predictions (FP), and false negative predictions (FN) are used as evaluation metrics.

### 3.2.1 NOD21 Detection

NODE21 [22] is released for MICCAI 2021 lung nodule detection challenge. This dataset has 4882 CXR scans with the ratio of patient: volunteer = 1134: 3748. It has 5524 annotations in total with maximum of three positive annotations for each scan. We split the overall images into training and validation sets at the ratio of 7:3 and meanwhile balanced out the proportion of patient and volunteer of each set. We run three different type of data input individually into Mask R-CNN network for 30,000 iterations with LR of 0.001, stochastic gradient descent optimizer (SGD), batch size of 2×4 on the $4 \times RTX\ A6000$ GPU cluster. Data Augmentation includes random scaling, random cropping, and random gaussian blur are implemented for robust training. As seen from **Table 1**, first, mix of raw and rib suppressed images in model training achieves the best detection scores compared to only train on raw CXRs or soft tissue images. Next, we evaluated the model trained with raw+soft-tissue CXRs on validation set with raw or soft tissue scans only, individually. And we can see that those two scenarios can roughly achieve the same level of performance with predictions made on raw CXR being slightly more competitive. Next, while we were making comparison between mix-trained and single-source-trained models, we found that mixed-trained detector achieves roughly 2-3% higher AUC score and meanwhile can locate relatively more nodes while significantly bring down the FP predictions than single-source-trained detector. Lastly, we found that there is no significant difference of model trained with single-source input except that detector trained with soft-tissue scans can slight reduce the number of FP predictions. **Figure 6** visually supports our quantitative findings with the first row showing that mixed-trained model can detect more nodules after suppression of ribcage the second row demonstrating that mixed-trained model can reduce FP predictions which are induced by rib shadows.

| Modality | Training Input | Validation Input | AUC | FN | FP | TP |
|---|---|---|---|---|---|---|
| Mask R-CNN | Raw | Raw | 94.76% | 48 | 1273 | 372 |
| | Soft tissue | Soft tissue | 95.32% | 45 | 1193 | 375 |
| | Raw + Soft tissue | Raw | 97.99% | 32 | 1070 | 388 |
| | | Soft tissue | 97.31% | 33 | 1082 | 387 |

**Table 1.** Evaluation results on NODE21 dataset. FN, FP, and TP show the absolute number of predictions for more straightforward comparison. The best performer is bold marked, and the lowest result is underlined.

### 3.2.2 ChestX-ray14 Classification and Localization

ChestX-ray14 [23] is a CXR set with text-mined fourteen lung diseases bounding box (bbox) annotations solely for the pre-split test set. This dataset is specially designed for weekly supervised or unsupervised learning tasks. ChestX-ray14 has in total 108,948 frontal view XR images and was pre-split into training and test sets prior release to the public but has only 984 images in the test set annotated with bboxes. To utilize this dataset for rib-suppression validation, we extract images with bboxes and then randomly sampled $3 \times 984$ healthy volunteers without annotations from the test set of ChestX-ray14 to prepare a dataset for supervised detection learning of common pulmonary diseases. Next, we split the sampled positive and negative scans into training and validation sets at the ratio of 7:3 and carefully balanced out the ratio of positive: negative = 1: 3. We run three different type of data input individually into Mask R-CNN network for 50,000 iterations with LR of 0.001, stochastic gradient descent optimizer (SGD), batch size of 2×4 on the $4 \times RTX\ A6000$ GPU cluster. Data Augmentation are implemented the same way as **Section 3.2.1**. As shown in **Table 2**, the statistics from ChestX-ray14 share similar findings as those from NODE21 in **Table 1**. Specifically, mix of raw and rib suppressed scans as training input achieves the best scores across the three different input combinations, has around 6-7% higher AUC than single-source-trained detector, and can largely reduce the number of FP predictions. Next, single-source-trained models roughly reach the same level of performance except that model trained with rib suppressed images makes less FP predictions. Moreover, since ChestX-ray14 is a dataset for multiple lung disease localization, we also present the AUC of each disease in **Table 3**. We found that though the scores vary among different diseases, we can draw the same conclusions as discussed for **Table 2**. **Figure 7** visually support our quantitative findings with ribcage the first row demonstrating that mixed-trained model can reduce FP predictions and the second row demonstrating that mixed-trained model can detect more abnormalities.

| Modality | Training Input | Validation Input | AUC | FN | FP | TP |
|---|---|---|---|---|---|---|
| Mask R-CNN | Raw | Raw | 80.54% | 137 | 3029 | 245 |
| | Soft tissue | Soft tissue | 81.55% | 138 | 2909 | 244 |
| | Raw + soft tissue | Raw | **87.16%** | 116 | 2644 | 275 |
| | | Soft tissue | 86.89% | 124 | 2701 | 267 |

**Table 2.** Evaluation results on ChestX-ray14 dataset. FN, FP, and TP show the absolute number of predictions for more straightforward comparison. The best performer is bold marked, and the lowest result is underlined.

| Input: Tr/Val  Disease | Raw / Raw | ST / ST | Raw + ST / Raw | Raw + ST / ST |
|---|---|---|---|---|
| **Atelectasis** | 79.21% | 77.35% | **85.27%** | 84.89% |
| **Cardiomegaly** | 83.47% | 82.03% | **93.84%** | 92.67% |
| **Effusion** | 84.98% | 84.72% | **91.02%** | **91.02%** |
| **Infiltration** | 69.84% | 70.00% | **75.21%** | 74.32% |
| **Mass** | 77.23% | 77.38% | **87.83%** | 85.79% |
| **Nodule** | 76.32% | 75.37% | **82.33%** | 81.29% |
| **Pneumonia** | 73.89% | 73.24% | **79.43%** | 78.33% |
| **Pneumothorax** | 82.43% | 82.32% | **91.37%** | 90.62% |
| **Consolidation** | 74.57% | 75.23% | **82.77%** | 81.98% |
| **Edema** | 88.14% | 87.28% | 95.32% | **95.33%** |
| **Emphysema** | 90.68% | 89.47% | **96.21%** | 95.79% |
| **Fibrosis** | 82.99% | 83.21% | **87.45%** | 86.76% |
| **Pleural Thickening** | 73.37% | 72.45% | **80.34%** | 79.85% |
| **Hernia** | 88.03% | 89.21% | **97.24%** | 96.79% |

**Table 3.** The AUC scores for 14 lung diseases on ChestX-ray14 dataset. Tr/Val represents training/Validation. The best results are bold marked, and the lowest ones are underlined.

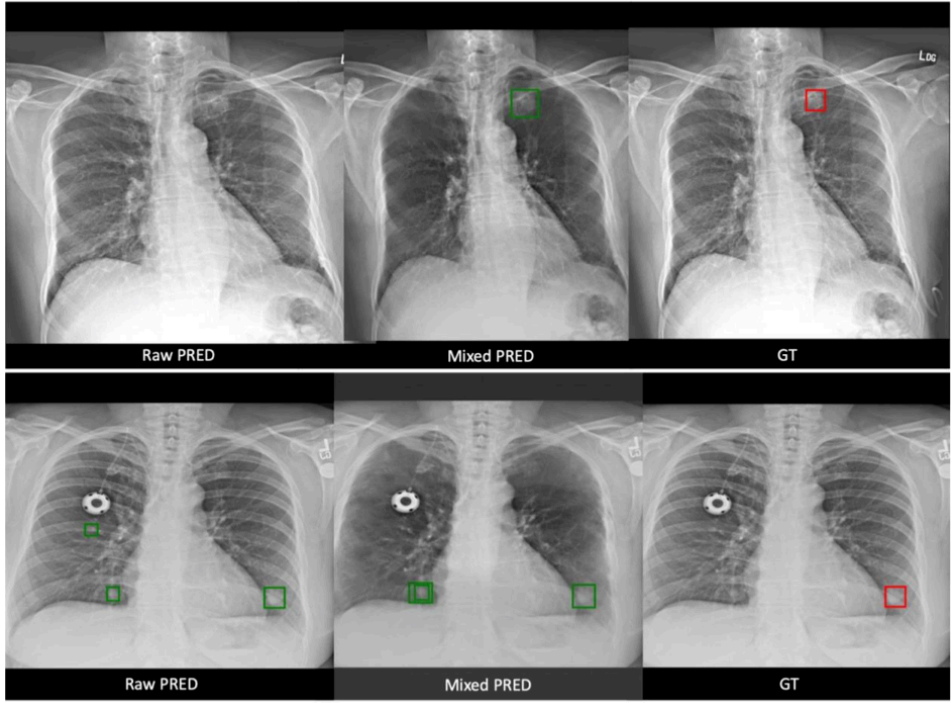

**Figure 6.** Two sample cases from NODE21 test set with each row presenting one patient. The figure is organized in three columns, having the first column showing predictions made from model trained with raw CXR only, the second column showing predictions made from mix-trained model, and the last column showing the GTs.

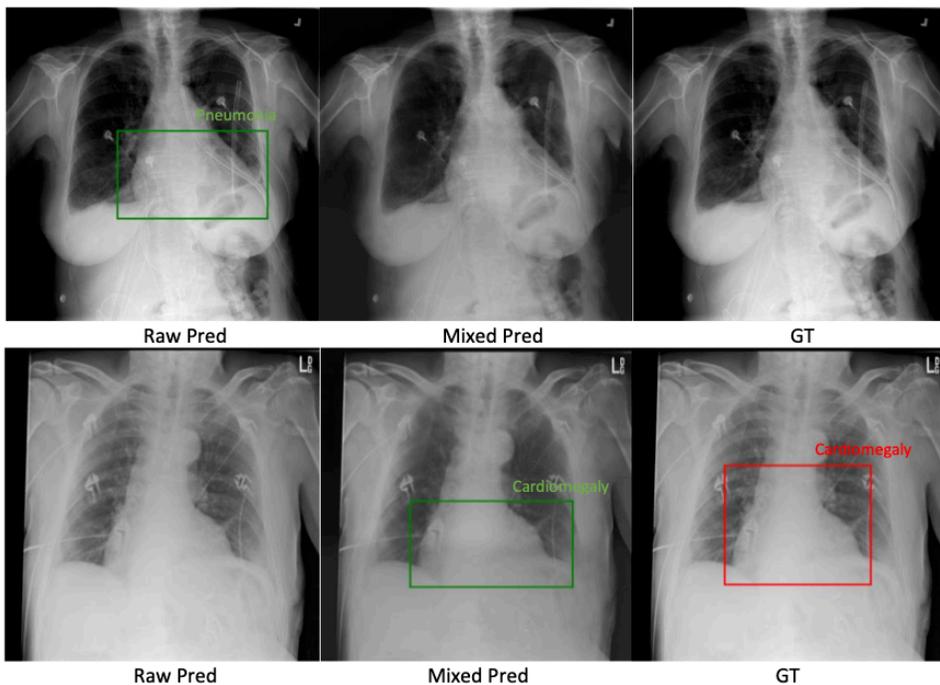

**Figure 7.** Two sample cases from ChestX-ray14 test set with each row presenting one patient. The figure is organized in three columns, having the first column showing predictions made from model trained with raw CXR only, the second column showing predictions made from mix-trained model, and the last column showing the GTs.

## 4. Discussion

In the current study, we first present the unsupervised ST-smoothing algorithm improved based on a physical model proposed by von Berg et. al [7] for CXR rib suppression. ST-smoothing can suppress high-contrast bony structures while maintaining the integrity of image quality, especially the contrast-detail level, as closely as to that of the original scan. Yet, ST-smoothing takes around an hour to process one CXR scan, requires patient-wise hyper-parameter tuning, and can only be applied on images paired with full annotated rib masks. The long running time and expensive labor cost in hyper-parameter tuning and rib annotation deviate from the clinical need of an efficient and automate solution. Hence, we consider using ST-smoothing to prepare a benchmark dataset having paired original and rib-suppressed CXR scans and then use that benchmark dataset to train a supervised learning algorithm which is fast and robust enough for practical clinical application. Followed by that consideration, we introduced a benchmark dataset named FX-RRCXR processed from VinDr-RibCXR [21] dataset, which in total has 245 paired raw and soft-tissue scans. Next, by seeing rib structures as noise spreading on soft-tissue images, we proposed a supervised image denoising network called SADXNet, trained on FX-RRCXR and able to predict a rib-suppressed image within a second. Lastly, we evaluate the quality of FX-RRCXR dataset and SADXNet in two downstream tasks including lung nodule detections using NODE21 [22] dataset and fourteen lung disease classification and localization using ChestXray14 [23] dataset.

In the past few years, big data has ushered in the possibility of a fundamentally different type of quantitative modelling. Instead of organizing hypothesis and then building statistical models to unfold a problem, people start to collect available data and then train DL models to map the question to its corresponding answer [13]. DL methods can be further classified into supervised and unsupervised learning, with supervised learning tending to be more advanced and accurate [28]. The critical edges of supervised deep network lie in its efficacy, automation and accuracy if having appropriate data for model training. Tons of existing works present that supervised DL can significantly outperform various types of conventional models in numerous fields [13] under the premise that decent training data is fed into network for adequate learning. Nonetheless, the scenario for constructing a supervised DL-based algorithm for CXR rib removal is never the same. Because preparing an appropriately paired CXR dataset for rib suppression is in an extremely particular setting, where the real GT – soft tissue CXR – is not theoretically acquirable. Consequently, the essential hurdle of supervised DL-based rib removal in CXR lies not only in proposing a superior network but more importantly in finding a suitable and large-enough training dataset. As mentioned, existing DL methods mainly use DRR or DE CXR modalities to prepare the learning set, which either have domain adaptation issue or are limited by the amount of available data [19]. Therefore, we start to break down this problem from a completely different angle. In specific, instead of setting conventional physical models as the benchmark for supervised DL model to beat, we should consider using neural network to offset the downsides of physical model – slow, intensive hyper-parameter tuning, and most requiring rib mask annotations. Therefore, we propose to use physical model to generate a qualitative dataset first and then train a supervised deep network based on that. As described in **Section 2.3.1**, it only takes up 100 epochs for SADXNet to converge and achieve almost 0 test RMSE, which further validate our conception that once gathering enough qualitative data, training a neural network to suppress ribs is not a difficult task. We will also release FX-RRCXR dataset upon the publication of this manuscript.

Generally, distributional shift is a major problem that may negatively affect the performance of deep learning model when we plan to put them into real clinical setting [29]. Since, compared to training data, test cases are most likely from distinct patient groups or collected from devices with different parameter setting, where some information was not exposed to DL models during training. However, we found that although SADXNet is trained on FX-RRCXR, it can still robustly suppress the rib structures on the scans from NODE21 and ChestX-ray14 datasets as shown in the middle column of **Figure 6** and **Figure 7**, which shows great promise of bringing our proposed workflow into real application.

Lastly, per downstream validation, both tasks show that solely training detection model on soft-tissue CXRs will not increase model performance compared to train on original scans. Whereas mixed training of two image sources can greatly reduce type I error while moderately increase the model sensitivity. We account the reasons in three-folds: 1) Why mixed training works better – We believe that it is the comparison of region of interest (ROI) with- and without covering from ribs that helps the learning process. 2) Why significant decrease in type I error – After assisting detector learning with soft-tissue scans, model is more likely to avoid recognizing the noise on rib structures, like edges, as ROI. 3) Why only moderate cut down of type II error - Since ribcage is not the only superimposed anatomy in CXR, suppression of rib structures on CXR can solely help model detect abnormalities covered or influenced by rib shadows. Other factors include heart, blood vessels, foreign bodies, and varied shapes of lung abnormalities are potential paths to explore for improvement of pulmonary disease diagnosis.

## 5. Conclusion

In this study, we improved ST-smoothing from von Berg et. al [7]. And based on that we introduced a paired dataset called FX-RRCXR that can potentially serve as benchmark dataset for supervised deep learning on CXR rib removal. Next, we proposed a denoising network called SADXNet which learns rib suppression via considering rib shadows as noise on top of soft-tissue scans. Lastly, we validated the rib removal results in two downstream tasks including lung nodule detection and common lung abnormities classification and localization. Experimental results from downstream tasks demonstrate the potential of FX-RRCXR dataset and SADXNet. In future study, CXR lung extraction assisted pulmonary abnormality study will be explore to a achieve theoretically noise-free and potentially more robust lung disease diagnosis system.

## 6. Acknowledgement

The study is supported in part by NIH R01CA259008 and DOD W81XWH2210044.